\documentclass{article}

\usepackage{arxiv}

\usepackage[utf8]{inputenc} % allow utf-8 input
\usepackage[T1]{fontenc}    % use 8-bit T1 fonts
\usepackage{hyperref}       % hyperlinks
\usepackage{url}            % simple URL typesetting
\usepackage{booktabs}       % professional-quality tables
\usepackage{amsfonts}       % blackboard math symbols
\usepackage{nicefrac}       % compact symbols for 1/2, etc.
\usepackage{microtype}      % microtypography
\usepackage{lipsum}

\usepackage{graphicx}
\usepackage{comment}
\usepackage{amsmath,amssymb} % define this before the line numbering.
\usepackage{color}
\usepackage{multirow}

\title{Robust Spatial-spread Deep Neural Image Watermarking}

\author{
  Marcin Plata$^{1,2}$ \\
  \texttt{marcin.plata@pwr.edu.pl}  \\
  \\
  $^{1}$ Department of Fundamentals of Computer Science \\
  Wroclaw University of Science and Technology \\
  Wroclaw, Poland \\
   \And
  Piotr Syga$^{1,2}$ \\
  \texttt{piotr.syga@pwr.edu.pl}  \\
  \\ 
  $^{2}$ Vestigit \\
  Wroclaw, Poland \\
  \\
}

\begin{document}
\maketitle

\begin{abstract}

Watermarking is an operation of embedding information into an image in a way that allows to identify ownership of the image despite applying some distortions on it. 
In this paper, we present a novel end-to-end solution for embedding and recovering the watermark in the digital image using convolutional neural networks.
We propose a spreading method of the message over the spatial domain of the image, hence reducing the \textit{local bits per pixel} capacity and significantly increasing robustness. 
To obtain the model we use adversarial training, apply noiser layers between the encoder and the decoder, and implement a precise JPEG approximation. Moreover, we broaden the spectrum of typically considered attacks on the watermark and we achieve high overall robustness, most notably against JPEG compression, Gaussian blur, subsampling or resizing.
We show that an application of some attacks could increase robustness against other \textit{non-seen during training} distortions across one group of attacks --- a proper grouping of the attacks according to their scope allows to achieve high general robustness. 

\end{abstract}

% keywords can be removed
\keywords{Blind watermarking, Robustness to attacks, Autoencoders, Neural networks, Spatial spreading}

\section{Introduction}
In the recent years the multimedia market has been steadily growing. An access to vast range of desired multimedia is provided in more convenient ways, e.g. Netflix offers offline access to movies and TV shows~\cite{NetflixOffline}. It also causes an increase in illegal redistribution of copyrighted content. One of the most efficient method to prevent such behaviour utilizes embedding of human-invisible watermark in a content. 
Watermarking  uses  the  fact  that a bandwidth of image is much higher that an amount of information which could be properly received and interpreted by human. It is well-known that a human eyesight is more sensitive to \textit{luminance} component of a color space than to \textit{chrominance}, i.e. one can recognize even small difference in a brightness of the image, but small color perturbations are oblivious to human's visual system. The watermarking is one among many properties operating on the surplus bandwidth that are used in such applications often alongside \textit{compression} or \textit{steganography}.  

In the watermarking model of communication, a user needs to embed a message into a digital image and send it to a recipient. Afterwards the image may be manipulated on by an attacker, however a legitimate user who shares a set of joint strategies of embedding and extracting the message should be able to recover the embedded message from the (possibly manipulated) image. The goal of the attacker is to modify the image, without significant deterioration, in order to destroy the embedded message, yet preserving the commercial value of the original data.   

During the work on watermarking techniques, we need to handle three following requirements \cite{survey}:
\begin{enumerate}
    \item \textit{transparency} concerns the quality of the image after the watermark encoding. In general, the original and watermarked images need to be perceptually similar. All distortions affected by the watermark embedding should be invisible for the human eyes, so that the value of the data for the consumers does not deteriorate. In our work, we utilized \textit{peak signal-to-noise ratio} (PSNR), which measures the pixel-wise difference between two images; 
    \item \textit{robustness} describes user's ability to decode the message from the encoded images after applying some signal processing operations on it. These operations could be applied intentionally, in order to destroy the watermark, or be a result of technical requirements or limitations. In this work, we used a terminology \textit{attacks} referring to these operations. Examples of attacks include cropping, resizing, Gaussian blur or JPEG compression;  
    \item \textit{capacity} was defined in \cite{2002} as "the number of bits a watermark encodes within a unit of time or work". In this paper, we additionally introduced \textit{local} or \textit{block bits per pixel capacity} to handle a limitation of convolutional layers. The size of the block could be delimited by calculating the longest distance on which information about any pixel is spread over the image using the encoder architecture based on the sequence of the convolutional layers, e.g. for one convolutional layer with the kernel size equal to \(7\), the block size is \(3\) and for two layers with kernel size equal to \(5\), it is \(4\). Note that, as opposite to steganography, we do not aim to embed the longest possible message, the main goal is to allow fitting essential information, as well as data needed for their correct retrieval, with possibly small changes of the covertext.
\end{enumerate}

In this paper, we introduce a novel technique of embedding a secret message into a digital image and extracting it using convolutional neural networks.
We proposed a method of spatial spreading of the secret message over the image, which significantly reduces the local (block) bits per pixel capacity, and at the same time retains the overall capacity of the image and preserves robustness on spatial attacks, such as rotating or cropping. Additionally, using the spatial spreading method significantly reduces the time of the training phase in comparison to previous solutions.
The proposed method has been validated against a wide group of attacks including lossy compression techniques, such as subsampling and JPEG compression, and spatial attacks, such as rotating and cropping. Despite considering such attacks by the multimedia community throughout the history of 'classic' watermarking, some of these attacks were neglected by other authors of recent watermark encoding solutions using neural networks, even though the attacks are easy to apply, and some of them are common components of a lossy compression techniques.

We also divide the considered attacks into five groups based on the scope they affect the image. Next, we show that  it is essential to apply the attacks from various groups in order to build a robust deep learning system for watermarking.
Finally, we evaluate the robustness of our method against the attacks in terms of the quality of the image  measured by \textit{peak signal-to-noise ratio} (PSNR). 

\textbf{Our contribution} is (1) a new architecture of the spatial-spread encoder and decoder as well as (2) the formulation of a loss function matching the architecture. (3) We improve the robustness against particular attacks in comparison to the current state-of-the-art methods, especially JPEG lossy compression algorithm, resizing and Gaussian blurring. (4) We handle new types of attacks, such as subsampling, which is a part of JPEG algorithm. (5) The resulting training framework required half the time in comparison to prior solutions. (6) We carry out the analysis of attacks' types -- we group the typical attacks according to their scope and show that an application of some attacks to the training pipeline could increase robustness against all distortions across a single group. (7) Our group-based analysis could be helpful in choosing the appropriate and balanced set of attacks applied to the noiser layers and deriving dependencies between them.

\section{Related work}

The problem of transparent and robust embedding of additional information into a digital domain was deeply studied for many years. Watermark solutions could be divided into two types \textit{non-blind} and \textit{blind}. The non-blind solutions require an original copy of the image for a detection step, whereas blind methods are able to detect a message encoded into the covertext without any additional data. Due to their easier application in real-life environment, most recent works has been focused on the blind approaches. Many solutions use spatial-to-frequency domain transformations, such as Discrete Fourier Transform (DFT)~\cite{4129062}, Discrete Wavelet Transform (DWT)~\cite{Najafi,7415839,Kumar2018ImprovedWI,makbol2014new,lai2010digital}, Discrete Cosine Transform (DCT)~\cite{5966517,Shivani2017RobustIE,patra2010novel} and others~\cite{8324041,8259288}. Extreme Machine Learning (EML) is another technique used for embedding watermarks into digital images which is gaining popularity over the last years~\cite{6252363,7415839,7966011}. Another method used widely for handling the watermark problem is Singular Value Decomposition (SVD) that was utilized in~\cite{makbol2014new,gupta2012robust,makbol2013robust,lai2010digital,loukhaoukha2011optimal} among others. Many presented works handled watermarking with combination of two or more techniques (e.g.~\cite{makbol2014new,7415839}). In recent years, we could also observe increased interest in applying deep learning methods into the watermarking field.  Authors of~\cite{Zhu_2018_ECCV} proposed a framework for training encoder and decoder networks in end-to-end manner due to adding noiser layers between the encoder and the decoder and an advisory network decided whether the images were encoded or not. A message was spread over all pixels on an image, hence allowing to achieve impressive robustness for cropping attacks. The paper was followed by~\cite{wen2019romark}, where the authors introduced a novel method of training the original architecture, called \textit{adversarial training}. They reported a high robustness against the attacks, however it resulted in  relatively low quality of encoded images measured by the PSNR. Another interesting approach for improving the robustness of a message detection was using an additional attack neural network for generating generic distortions introduced in~\cite{luo2020distortion}. The authors of \cite{zhong2019robust} designed a fully automated deep learning-based system for watermark extraction from camera-captured images. In \cite{ZeroCNN}, the authors used convolutional neural networks for zero-watermarking which does not modify the image but extracts some characteristics from the image in order to linking it with an owner. The paper \cite{JpegRotate} described a deep learning solution robust against JPEG compression and rotating. In RedMark \cite{ReDMark}, there was a special transform layer used on an image before feed forwarding the encoding neural network and they worked out a differentiable approximation of JPEG. In \cite{Approx2019}, authors also proposed a method of JPEG approximation.

\section{Method}
\subsection{Formulation}
The main goal of the watermarking method is encoding additional information, called \textit{watermark}, into a digital image, called \textit{cover image} in a way that allows recovering the watermark by a legitimate user. The watermark needs to be robust against some signal processing operations, called \textit{attacks}. In this work, we considered the following attacks: cropping, cropout, dropout, rotation, Gaussian smoothing, subsampling 4:2:0, JPEG compression, resizing. All attacks as well as the watermark encoding need to ensure the transparency.

We aim to encode a binary message \(m \in \{0,1\}^L\), where \(L \in \mathbb{N}_+\), in the cover image \(I_{c}\) of shape \((H \times W \times \mathrm{Ch}) \in  \mathbb{N}_{+}^{3}\). The result of this operation is the encoded image \(I_{e}\) containing the hidden watermark \(m\). Both images \(I_c\) and \(I_e\) need to be perceptually indistinguishable. Next, an attacker distorts \(I_{e}\) by applying selected attacks in order to prevent the extraction of \(m\) from the encoded image. An output after distortions is a \textit{noised image} \(I_{a}\) which has three channels and unspecified width and height. Finally, we try to extract a hidden message \(m' \in \{0,1\}^L\) from \(I_a\) that satisfies \(\parallel m - m' \parallel < \delta\).

\subsection{Architecture}

\begin{figure*}
    \centering
    \includegraphics[width=\textwidth]{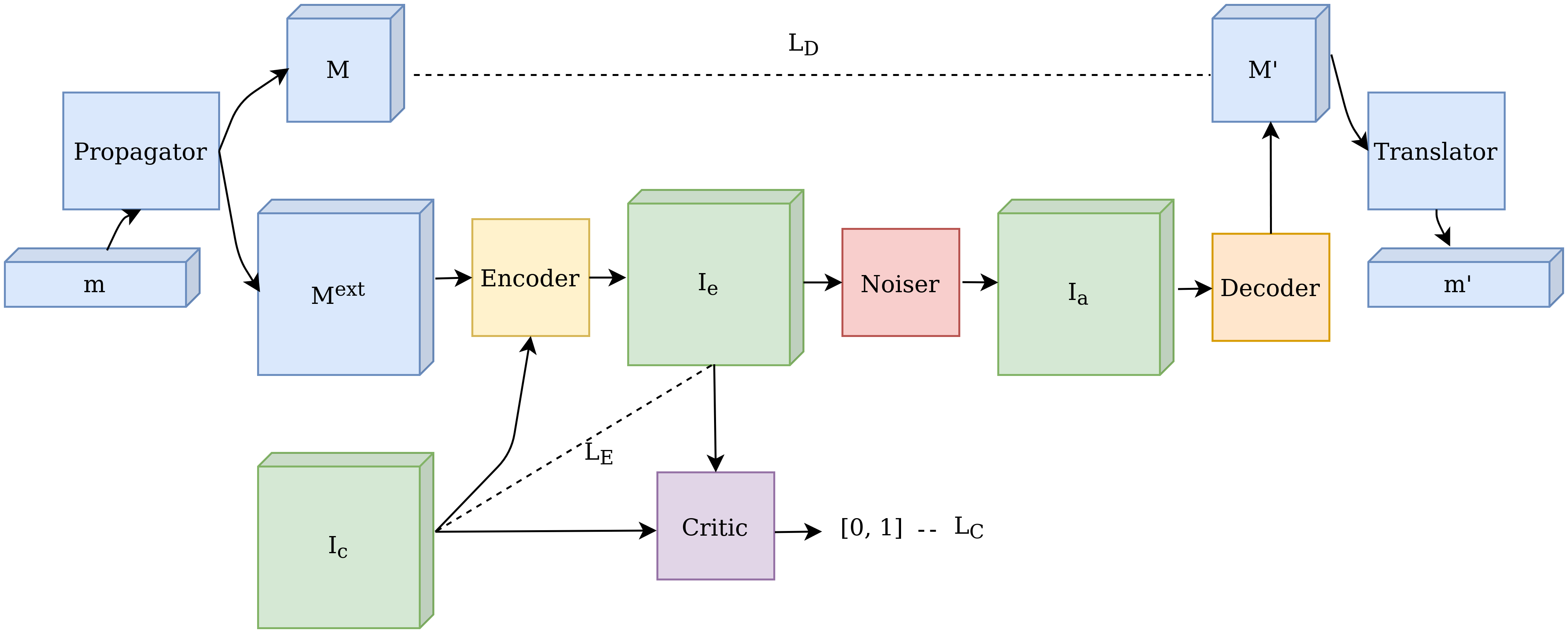}
    \caption{The sketch of a training pipeline. The propagator converts a message \(m\) in two ways - \(M^{ext}\) which is pushed through the training pipeline and \(M\) used to calculate the loss \(L_D\). The encoder encodes \(M^{ext}\) into an image \(I_c\) and returns a encoded image \(I_e\). The noiser distorts \(I_e\) in order to mimic attacks and expose possible ways of distortions to the neural networks. The decoder gets a distorted image \(I_a\) and extracts an encoded message \(M^{'}\) in a shape of \(M\). Finally, the translator calculates a final message \(m'\) based on \(M^{'}\). The critic is an adversarial training component used to improve a quality of \(I_e\).}
    \label{fig:arch}
\end{figure*}

The architecture proposed in the paper consists of six  main components. Three of them are trainable neural networks called \textit{encoder} \(E_{\phi}\), \textit{decoder} \(D_{\gamma}\) and \textit{adversarial critic} \(C_{\omega}\), where \(\phi\), \(\gamma\), \(\omega\) are trainable parameters. An additional component is \textit{noiser} \(A\) used for performing attacks on the encoded image. We also specified two deterministic algorithms called \textit{message propagator} \(P\)  and \textit{message translator} \(T\). The overall sketch of the architecture was presented in Figure~\ref{fig:arch}. 

We denote a \(i\)-th bit of the message \(m\) as \(m_{i}\) and we represent the message using a sequence of tuples, where the tuple \(s_i = (i, m_{ki},  m_{ki+1} \dots, m_{ki+k-1}) \), where  \(i \in \{0, 1, \dots, \frac{L}{k}-1\} \) and \(1 \leq k \leq L\). In particular, for \(k = 1\), we are able to represent the message as a trivial sequence of tuples \((i, m_i)\) for \(i \in \{0, 1, \dots, L-1\}\). We also define a function \(bin_n: \mathbb{N}  \to  \{0, 1\}^n\) which for a given value returns its binary representation of a length equal to \(n\). \(b \in \mathbb{N}_{+}\) defines a block size containing replicated tuples. 

The propagator \(P_{nkb}: \{0, 1\}^{L} \to \{0,1\}^{\frac{H}{b} \times \frac{W}{b} \times (n + k)}\) is a function which executes following steps:
\begin{enumerate}
    \item convert the message \(m\) into a sequence of tuples \((s_0, s_1, \dots, s_{\frac{L}{k}-1})\),
    \item for every \(i\), convert the first element of a tuple \(s_{i}\) to the binary representation \(bin_n(s_{i0})\), 
    flatten the tuple \(s_i\), and unsqueeze to \(s_i  \in \{0, 1\}^{1 \times 1 \times (n+k)}\),
    \item build a spatial message \(M \in \{0, 1\}^{\frac{H}{b} \times \frac{W}{b} \times (n+k)}\) by randomly assigning tuples \(s_i \) to slices \(M_{xy}\), where \(x \in \{0, \dots, \frac{H}{b}-1\}\) and \(y \in \{0, \dots, \frac{W}{b}-1\}\). Note, we allow a production of redundant data in \(M\), i.e., inserting more that one tuple \(s_i\).
\end{enumerate}
We also need to extend \(M\) if the message is an input to the encoder. In such case one additional step is made: 
\begin{enumerate}
    \setcounter{enumi}{3}
    \item every splice \(M_{xy}\) is replicated $b$ times in horizontal and vertical direction (namely, the slice \(M_{xy} \in \{0, 1\}^{1 \times 1 \times (n+k)} \) is converting to \(M_{xy} \in \{0, 1\}^{b \times b \times (n+k)} \)).
\end{enumerate}

\begin{figure*}
    \centering
    \includegraphics[width=\textwidth]{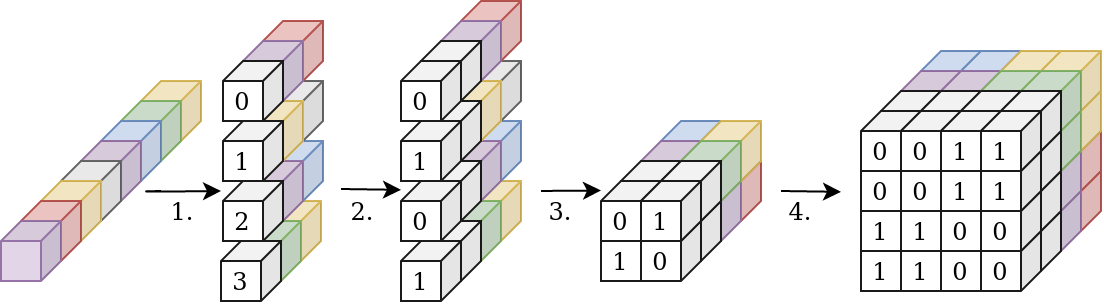}
    \caption{The visualization of steps of the propagator \(P_{nkb}^{ext}\) for parameters \(n=2\), \(k=2\), \(b=2\), \(L=8\), \(W=4\) and \(H=4\). The numbers under the arrows refer to the propagators steps.}
    \label{fig:prop_vis}
\end{figure*}

If the additional step needs to be executed, we denote the propagator by \(P_{nkb}^{ext}\) and achieve \(M^{ext} \in \{0,1\}^{H \times W \times (n + k)}\). The visualization of the propagator is presented in Figure~\ref{fig:prop_vis}.

The output of the propagator  \(P_{nkb}^{ext}\) together with the cover image \(I_{c}\) is used by the encoder \(E_{\phi}\) to produce the encoded image \(I_{e}\), i.e.:
\begin{align}
I_{e} = E_{\phi}(I_{c}, M^{ext}). 
\end{align}
We follow by applying the attacks on the image \(I_e\) by:
\begin{align}
I_{a} = A(I_{e}, I_{c}, M^{ext}). 
\end{align}
Note that, some attacks required the cover image \(I_c\), e.g. dropout. For the crop attack, we also cropped the message \(M\) during the training. The decoder \(D_{\gamma}\) tries to extract the message \(M\) having an access only to \(I_{a}\):
\begin{align}
M' = D_{\gamma}(I_{a}) \in \{0, 1\}^{\frac{H}{b} \times \frac{W}{b} \times (n+k)}. 
\end{align}
Additionally, we use \(C_{\omega}\) to rate if \(I_e\)  is similar to \(I_{c}\), i.e., whether the watermarked image is of an acceptable quality for end-users:
\begin{align}
 C_{\omega}(I \in \{I_e, I_c\}) \in [0,1]. 
\end{align}
The last element of the architecture is the message translator \(T_o\).  It is a deterministic function which calculates the final message \(m'\) based on the decoded message \(M'\). The process of the calculation is similar to the k-Nearest Neighbours algorithm. For every \(i \in \{0, 1, \dots, \frac{L}{k}-1\} \), we find \(o\) tuples from \(M'\) with first \(n\) values (referred by binary index) that are closest to \(bin_n(i)\), i.e., we choose a tuple with coordinates \(xy\) if \( || bin_n(i) - M'_{xy[0, \dots, n-1]} ||_2 \) is one of \(o\) lowest values. Then, we calculate mean values for each element encoding a bit of the message, i.e. elements from the tuple on positions \((n, \dots, n+k-1)\), enabling us to predict all bits from  \(m'\).

The whole architecture allows to encode the message \(m\) in the cover image \(I_c\) and reduce a number of the local (block) bits per pixel capacity. 
The state-of-the-art and recent architectures of encoders \cite{Zhu_2018_ECCV,wen2019romark,luo2020distortion} are based on the convolutional layers. It means that the encoder embeds the message locally, without an access to the whole image. This architecture of the encoder provokes two ways of encoding the message. (1) Encoding only subset of the whole message depending on the pixels color space, e.g. encode some bits only if a tone of the pixel is close to blue. This way of encoding is risky and unreliable. (2) Attempting to encode the whole message locally (in the block of pixels). A results' analysis of the robustness on attacks, in particular, the high accuracy against cropping attack, indicated that the second way of the message encoding is more likely. Thus, we proposed the solution for reducing the local bits per pixel capacity and improved the robustness against attacks, especially smoothing-type attacks. 

The proposed architecture spreads fractions of the message \(m\) over the image in the form of tuples \( s_i = (i, m_{ki},  m_{ki+1} \dots, m_{ki+k-1}) \), where \(i \in \{0, 1, \dots, \frac{L}{k}-1\} \) and \(1 \leq k \leq L\). Note that the spread is performed in a block fashion rather than assigning the whole message \(m\) to every single pixel. For example, we could encode the message of length \(L = 32\) by splitting it into 8 patches of length equal to 4 (\(k=4\) and \(n=3\)). Thus, we are able to encode the patch by 7 bits, where we need 3 bits for the index of the patch and 4 bits for the corresponding fraction of the massage. %The analysis of other cases was presented in Figure~\ref{fig:kn}. 
During our experiments, we achieved the best results for \(k=2\).

\subsection{Loss functions}

We formulated a novel loss function for training our models using gradient descent algorithm. Our general objective contains three separated loss functions \(L_E\), \(L_D\) and \(L_C\), for training the encoder \(E_{\phi}\), the decoder \(D_{\gamma}\) and the critic \(C_{\omega}\), respectively. The nosier \(A\) are inside the training pipeline and do not contain training parameters. Furthermore, the message propagator \(P_{nkb}\) and translator \(T_o\) are deterministic algorithms outside of the training pipeline.

The aim of the loss function \(L_E\) is keeping images \(I_c\) and \(I_e\) similar. It was formulated as follow:
\begin{align}
L_{E}(I_c, I_e) = \mathrm{MSE}(I_c, I_e) = \frac{1}{H \cdot W \cdot \mathrm{Ch}} || I_c - I_e ||_2^2,
% \frac{1}{HWC} \sum_{h=0}^{H-1} \sum_{w=0}^{W-1} \sum_{c=0}^{C-1} (I_{c_{hwc}} - I_{e_{hwc}})^2, 
\end{align}
where \(\mathrm{MSE}\) is a standard Mean Square Error function. The loss function \(L_{D}\) works on the similarity between propagated messages \(M\) and \(M'\). However, as \(M\) contains redundant data, i.e. the same tuples, we do not need to perfectly recover the message. Our aim was to extract a subset of tuples with "high confidence" of information. Thus, we formulated the loss function \(L_D\) as a combination of mean and variance functions:
\begin{align}
L_{D}^{mean}(M, M') & = \frac{b^2}{H \cdot W} \sum_{h=0}^{H_b}  \sum_{w=0}^{W_b} Mean(| M_{hw} - M'_{hw} |) \; \\
& =  \frac{b^2}{H \cdot W \cdot (n+k)} || M - M' ||_1
\end{align}
and 
\begin{align}
L_{D}^{var}(M, M') = \frac{b^2}{H \cdot W} \sum_{h=0}^{H_b}  \sum_{w=0}^{W_b} Var(| M_{hw} - M'_{hw} |),
\end{align}  
where \(H_b = \frac{H}{b}-1\) and \(W_b = \frac{W}{b}-1\) and the operator \(|\cdot|\) returns the absolute value of every element of the vector. The final loss function is \(L_D = \lambda_D^{mean} L_D^{mean} + \lambda_D^{var} L_D^{var}\). Such formulation of the loss function promotes learning of \textit{all elements in some tuples} over \textit{some elements over all tuples}.

We also defined an adversarial training for the encoder \(E_{\phi}\) and the critic \(C_{\omega}\), so that better visual similarity of the images \(I_c\) and \(I_e\) was achieved. For the encoder \(E_{\phi}\), we expected to produce images following the transparency requirement, thus we defined the loss function \(L_C^E = log(1 - C_{\omega}(I_e))\). On the other hand, the role of the critic  \(C_{\omega}\) was to distinguish between the "real" images \(I_c\) and the modified image \(I_e\), thus in this case we defined the loss function  \(L_C^C = log(1 - C_{\omega}(I_c)) + log(C_{\omega}(I_e))\).

Finally, we ran gradient decent algorithm on \(\phi\) and \(\gamma\) parameters in order to minimize the loss function over the distribution of images \(I_c\) and messages~\(M\):
\begin{align}
\mathbb{E}_{I_c, M}[ \lambda_E L_E + \lambda_D^{mean} L_D^{mean} + \lambda_D^{var} L_D^{var} + \lambda_C L_C^E],
\end{align}  
where \(\lambda\)-s are weights for particular losses. We simultaneously conducted a training of \(C_{\omega}\) with to minimize the loss function over the distribution of images \(I_c\) with respect to $\omega$:
$\mathbb{E}_{I_c}[L_C^C]$.

%\subsection{Networks' architectures}
\subsection{The architecture of the networks}

The main block applied to the neural networks, i.e. the encoder \(E_{\phi}\), the decoder~\(D_{\gamma}\) and the critic \(C_{\omega}\), is a sequential structure of a convolutional layer with 64 channels, the kernel size equal to \(3 \times 3\), stride equal to \(1 \times 1\) and padding equal to \(1 \times 1\), utilizing batch normalization layer and ReLU activation. All networks operate on images in YCbCr color space. %, the encoder returns images in the same colorspace.

The encoder \(E_{\phi}\) contains five sequential blocks, where the first block is fed by the concatenated tensor of the image  \(I_c\) and the spread message \(M^{ext}\). Next, the tensor \([I_c, M^{ext}] \) is also concatenated with the input before every second convolutional layer, i.e., 1$^{\text{st}}$, 3$^{\text{rd}}$ and 5$^{\text{th}}$ layer has an access to the cover image and the spread message. The last encoder layer is a convolution with 3 channels and default parameter values. Note, that the number of layers in the encoder \(E_{\phi}\) does not exceed the other state-of-the-art methods, e.g. \cite{Zhu_2018_ECCV,wen2019romark,luo2020distortion,ReDMark}. It is important in the context of a time efficiency as in many practical scenarios (e.g. streaming) the encoder needs to work in real-time.

The decoder \(D_{\gamma}\) takes an encoded image \(I_e\) and puts it through 6 sequential blocks. Then, we apply an adaptive average pooling layer which produces a tensor with size equal to \(\frac{H}{b} \times \frac{W}{b} \times 64\). Next, the tensor is fed to the sequential block with 64 channels, the kernel size and the padding equal to \(1 \times 1\). The last layer is the separated convolution layer with \(k+n\) channels, the kernel size and the padding remain unchanged. Thus, the decoder returns a tensor with the same size as \(M\). The last two convolutional layers imitate fully connected layers for every spatial element of the output over channels. Note that during our experiments we did not change the size of the tensor produced by the adaptive pooling, i.e. the decoder returned the output tensor with the same size also after cropping or resizing attacks. Executing actions regarding attacks' types could improve the robustness of the method, but requires a method to recognize the attack's type and counters the end-to-end approach, thus we decided to return \(M'\) with the same size in every case.

The critic \(C_{\omega}\) consists of three sequential blocks, an adaptive average pooling layer which produces a 64-dimensional vector, then a fully connected layer. The critic returns the value describing a similarity of the input image to real images. 

In our experiments we also considered the models' architecture used in \cite{Zhu_2018_ECCV,lai2010digital}. In this scenario, we did not change the architectures of the encoder and the discriminator, while we needed to modify the last layers of the detector to handle our spatial-spreading method. We replaced a global average pooling with the adaptive average pooling and used the same sequence of layers as in our previously described architecture.
  
\subsection{Noiser layers and Attacks}

We selected some nosier layers which we later applied during the training process. We exposed to the neural networks various kinds of distortions which they needed to handle in order to increase the performance. By this, we were able to determine a way of training of the neural networks. The types of selected distortions included cropping and cropout, dropout, Gaussian smoothing, rotation, subsampling 4:2:0, approximation of JPEG and resizing.

The crop distortion returns a cropped square of the image \(I_e\) of a specified area ratio \(p = \frac{H^{new}W^{new}}{HW}\). The cropout attack works similar to the crop, it crops the square of the image \(I_e\) and instead of discarding the rest of the image, it replaces the outer area by the image \(I_c\). As in \cite{Zhu_2018_ECCV}, we decided to use the image \(I_c\) as the background for the encoded image \(I_e\) as this simulates a binary symmetric channel (BSC), which is a standard model considered in information theory,
where a receiver does not have knowledge if the obtained bit is correct or wrong. %Applying monotonic or random color of pixels of the extant area imitates other simple communication model --- binary erasure channel (BEC). 
The cropout attack was parameterized by us with a value \(p\) equal to a ratio of the cropped area over the entire image area. The dropout attacks keeps a percentage \(p\) of the pixels of the image \(I_e\) and the rest pixels replaces with corresponding pixels of the image \(I_c\). As in the cropout, this procedure also simulates the BSC model. Gaussian smoothing was done with a parameter \(\sigma\) (a kernel width). Note, that even though it is similar attack to using median filter, it is far less probable, as the resulting image is of higher quality, hence of a higher worth to the end-user (hence to the copyright violator). 

Next four attacks are our extension of those presented in~\cite{Zhu_2018_ECCV,wen2019romark}. The rotation attack rotates the image by \(\alpha\) degrees. The subsampling 4:2:0 is applied in many digital compression algorithms, such as JPEG or MPEG, and is the most popular from chroma-subsampling variants (e.g. 4:2:2, 4:1:1). It reduces the image channels \textit{Cb} and \textit{Cr} by calculating an average value of every square of \(2 \times 2\). The procedure could be done using a 2D convolutional layer with one channel, kernel size equal to \(2 \times 2\), stride equal to \(2 \times 2\) and weights set to \(0.25\). We also used a resize attack with a scale factor \(s = \frac{H^{new}}{H} = \frac{W^{new}}{W} \). We handled two types of interpolation -- \textit{Nearest neighbours} and \textit{Lanczos}. 

\subsection{Approximation of JPEG}
Lossy compression algorithms could be considered as most efficient attacks against a wide range of watermarking protocols. This comes from the fact that algorithms such as JPEG are very efficient in removing barely visible objects and information which is not essential for the viewer. On the other hand, all watermarking techniques aim at changing the image in a way that is hardly noticeable for the viewer and, later, to retrieve it.  Thus, it was necessary to apply compression in the training pipeline, in order to obtain an appropriate design for the encoder and the decoder training. The main inconvenience of the JPEG is a rounding operation applied on quantized frequency-domain elements of the image. The derivative of the round function is indeterminate for points \(x \in \mathbb{Z}\) and equal to \(0\) in the rest of the domain. Thus, using the rounding function in the middle of the training pipeline is impossible due to halting the update of the neural networks weights by the gradient descent algorithm. Although there is a method of approximating the compression~\cite{Approx2019}, in order to use it for a subsampling attack training a different approach had to be made. 
We proposed an approximation of JPEG compression which executes the following steps for the image \(I\): (1) converting to YCbCr color space, (2) subsampling 4:2:0, (3) splitting separately every channel into blocks of \(8 \times 8\) (4) applying the Discrete Cosine Transform (DCT), (5) dividing by the quantization table \(Q\) and (6) applying the approximation of the rounding. The last two steps, we formulated as follows:
\begin{align}
I'_{ij} = \begin{cases}
  0, & \text{if } -\frac{1}{2} \leq \frac{I_{ij}}{Q_{ij}} \leq \frac{1}{2}, \\
  \frac{I_{ij}}{Q_{ij}} + \delta Q_{ij}, & \text{otherwise},
\end{cases}
\end{align}  
where \(\delta \sim N(0, \sigma^2)\), \(I_{ij}\) is the frequency-domain element of the image and \(Q_{ij}\) is the related element of the quantization table. For our experiments, we set \(\sigma = 0.01\). We used the standard quantization table for the quality parameter \(q = 50\) and we modified the elements of the table \(Q\) for different \(q\) in accordance with the JPEG standard \cite{1993_jpeg_still,PARKER2017}. 
%To the best of our knowledge it is the most precise differentiable approximation of the classic JPEG applied to the deep learning training pipeline for the watermarking. 
For the evaluation procedure, we used the standard JPEG.

\subsection{Training details.}
The method was trained on the COCO dataset \cite{coco}. We used 10000 randomly-sampled cover images for the training subset and 1000 for the validation subset. Both subsets were disjoint. Both the messages and the spatial spreading was chosen at random. The parameters \(\lambda_E\), \(\lambda^{mean}_D\), \(\lambda^{var}_D\) and \(\lambda_C\) were set to \(4.0\), \(1.0\), \(1.0\) and \(0.01\), respectively. We used Adam~\cite{adam} with learning rate equal to \(0.001\) (other parameters had default values) for the stochastic gradient descent optimization. The models were trained with batch size equal to \(12\). The final training with applied all nosier layers took 100 epochs.

\section{Analysis of the attacks}

We observed that most of the attacks considered by us could be assigned into more general groups based on their specific characteristics. Thus, we classified attacks regarding the way in which they affect the image. We also assumed that after any attack a content of the image needs to be visible and its quality has to be acceptable to customers. With these assumptions, we 
specified five types of attacks:
\begin{itemize}
  \item \textbf{Pixel-specific}, where we modify only single pixels (without considering any others) by changing color, adding noise, replacing pixels by other random ones, removing some pixels or changing their position on the image. In this group we could specify two subgroups: one that applies \textit{one modification on all pixels}, and the other that applies \textit{one modification on a subset of pixels}.
  A characteristic of this group is that we have an access to a smaller subset of non-modified pixels after attacks or all pixels were transformed in the same specific way. To this group, we selected some attacks such as color space conversions, cropping, cropout, dropout and rotation.
  \item \textbf{Local}, where we modify pixels with regard to their neighborhoods. In this group, all pixels are modified during attacks, but only neighbours of the pixels affect the results (e.g., subsampling, Gaussian blur and resizing).
  \item \textbf{Domain}, where modifications are domain-specific and even small changes in limited neighbourhoods could affect globally on an image represented in a different domain. This group of attacks includes all transform methods, e.g. Discrete Cosine Transform (DCT).
  \item \textbf{Mixed}, where a final modification is a combination of methods from other groups. Here we could distinguish JPEG which combines color space conversion, subsampling and locally applied DCT.
\end{itemize}
The analysis of attack types could be important and helpful in the context of designing the training pipeline. Most of the recent deep learning solutions for watermarking use additional noiser layers in order to improve robustness for particular attacks (e.g. \cite{Zhu_2018_ECCV,wen2019romark,ReDMark}). It requires selecting a finite set of attacks applied during the training process. Moreover, all attacks in the training pipeline need to be differentiable as the noiser layers are usually embedded before the neural network responsible for the message's detection. As such, it requires deferential approximations of non-differentiable attacks, e.g. \textit{JPEG} compression. An appropriate choice of attacks for a training pipeline could cause a high robustness for other attacks which were not applied to the training pipeline.  In~\cite{luo2020distortion}, where authors proposed a distortion agnostic method using adversarial neural networks, we could observe that even small perturbations generated by attacks classified by us into the \textit{local} group noticeably decrease an accuracy of the message detection. It implies that the neural network generated distortions belonging to the \textit{pixel-specific} or \textit{domain} groups and ignored attacks similar to these from the \textit{local} group. In our work, we focused on selecting a special set of attacks which covers all four groups. 
\begin{figure*}
    \centering
    \includegraphics[width=\textwidth]{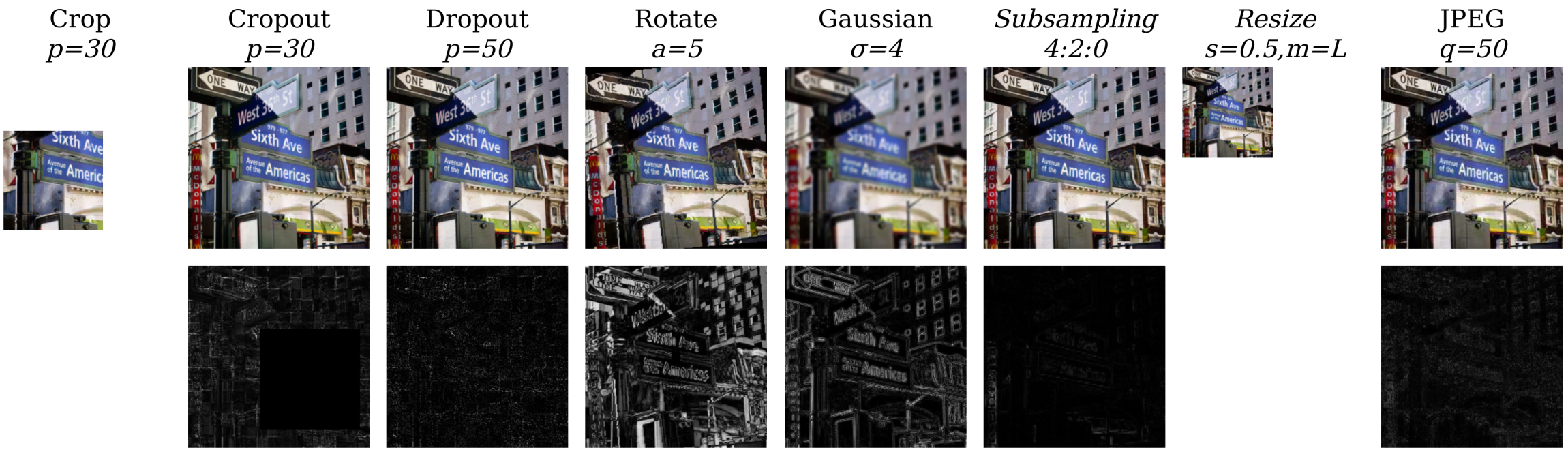}
    \caption{The visualisation of attacks' applications. The row above refers to the noised image \(I_a\) and the row below refer to the normalized difference between the noised image \(I_a\) and the encoded image \(I_e\). We used min-max normalization.}
    \label{fig:atc}
\end{figure*}

\paragraph{Robustness on exclusionary attacks' selection.}
We conducted an experiment with training the pipeline with only a subset of the attacks chosen from only one of the mentioned groups, and we observed its impact on the robustness against attacks from the same group and other groups. The results were presented in Table~\ref{table:exp}. The experiments confirmed that there exist a correlation between the ways of image modification by particular attacks and stronger correlations are noticeable between attacks belonging to the same group. It is trivial to notice, that crop and cropout attacks do not modify a whole patch of an image, i.e. the decoder has an access to the non-modified patch. The dropout attack changes random pixels, but still the decoder could detect the message based on not-modified pixels. Thus, applying only the subset of the attacks during training  achieves more general robustness on a wider collection of attacks from the same group.

\setlength{\tabcolsep}{4pt}
\begin{table}
    \begin{center}
    \caption{The results of the experiment of applying attacks from the same group during the training process. The values in the table refers to the bit accuracies. The red color indicates the attacks which were used during the training and the blue color refers to the best accuracy achieved for the non-applied attacks. Note, that best results were achieved around the same groups of attacks.}
    \label{table:exp}
    \begin{tabular}{c c c c }
    \hline\noalign{\smallskip}
    \multirow{2}{*}{Attacks} & \multicolumn{3}{c}{Noiser Layers} \\
    \cline{2-4}\noalign{\smallskip}
     & Identity & \shortstack{Crop(\([0.3, 0.9]\)) \\ Dropout(\([0.3, 0.9]\))} & \shortstack{Gaussian(\(\{3,5\}\)) \\ Subsampling(4:2:0)} \\
    \hline\noalign{\smallskip}
    Identity & \textcolor{red}{0.999} & \textcolor{red}{0.991} & \textcolor{red}{0.985} \\
    \hline\noalign{\smallskip}
    Crop(p=$0.3$) & 0.847 & \textcolor{red}{0.894} & 0.833 \\
    Cropout(p=$0.3$) & 0.793 & \textcolor{blue}{0.875} & 0.672 \\
    Dropout(p=$0.5$) & 0.530 & \textcolor{red}{0.972} & 0.574 \\
    Rotate($\alpha$=$5^{\circ}$) & 0.754 & \textcolor{blue}{0.821} & 0.780 \\
    \hline\noalign{\smallskip}
    Gaussian(\(\sigma\)=$5$) & 0.823 & 0.564 & \textcolor{red}{0.981} \\
    Subsampling(4:2:0) & 0.524 & 0.623 & \textcolor{red}{0.980} \\
    Resize(s=$0.5$, m=$L$) & 0.511 & 0.532 & \textcolor{blue}{0.735} \\
    JPEG(q=$95$) & 0.502 & 0.512 & \textcolor{blue}{0.783} \\
    \hline\noalign{\smallskip}
    \end{tabular}
    \end{center}
\end{table}
\setlength{\tabcolsep}{1.4pt}

\section{Watermark robustness}

In this section, we presented the evaluation of our method and the comparison with the current state-of-the-art solutions. The experiments were done for the images of the size \(256 \times 256\) and the message of the length \(L = 32\). Our main goal was reducing the local bits per pixels capacity, thus we set \(k=2\). By this, the number of bits required for storing the patch (tuple) was equal to \(6\) and the number of the patches was equal to \(16\). The tuple stored two bits of the message and the related index which took four bits. The block size \(b\) was set to \(16\). In order to spread all patches over the image, we needed to locate \(16\) blocks with the size equal to \(16 \times 16\) pixels. It indicated that the smallest size of the image was equal to \(64 \times 64\) pixels. The final method was trained with all types applied to the noiser layers. We considered the bit accuracy as a metric of the robustness against attacks. The results of the robustness on attacks were presented in Table~\ref{table:res}.

\setlength{\tabcolsep}{4pt}
\begin{table*}
    \begin{center}
    \caption{The results of the bit accuracy for selected attack types and the comparison with the state-of-the-art methods. The results in the column \textit{Spatial+Concat} were achieved using the spatial-spreading method and the encoder architecture with concatenation of $[I_c, M^{ext}]$ with every second convolutional layer, while in the column \textit{Spatial}, we provided the results for the standard encoder architecture used in \cite{Zhu_2018_ECCV,lai2010digital}. The evaluation was provided for the capacity equal to 32 bits. Note that, the resizing modes were not specified in \cite{luo2020distortion} and \cite{ReDMark}.}
    \label{table:res}
    \begin{tabular}{c c c c c c }
    \hline\noalign{\smallskip}
    \multirow{2}{*}{Attacks} & \multicolumn{4}{c}{Methods} \\
    \cline{2-6}\noalign{\smallskip}
     & \textit{Spatial+Concat} & \textit{Spatial} & HiDDeN \cite{Zhu_2018_ECCV} & DADW \cite{luo2020distortion} & RedMark \cite{ReDMark} \\
    \hline\noalign{\smallskip}
    Identity & 1.000 & 1.000 & 1.000 & 1.000 & 1.000 \\
    \hline\noalign{\smallskip}
    Crop($p$=$0.3$) & 0.832 & 0.883 & \textbf{1.00} & \textbf{1.00} & - \\
    Cropout($p$=$0.3$) & 0.902 & 0.901 & \textbf{0.940} & - & 0.925 \\
    Dropout($p$=$0.5$) & 0.962 & 0.889 & \textbf{1.0} & \textbf{1.0} &  \(\approx\)0.990 \\
    Rotate($\alpha$=$5^{\circ}$) & \textbf{0.842} & 0.828 & - & - & - \\
    \hline\noalign{\smallskip}
    Gaussian($\sigma$=$2$) & \textbf{0.986} & 0.982 & 0.960 & 0.600 & 0.500 \\
    Gaussian($\sigma$=$4$) &\textbf{0.982} & 0.980 & 0.820 & 0.500 & 0.500 \\
    Subsampling(4:2:0) & \textbf{0.984} & 0.980 & - & - & - \\
    Resize(s=$0.5$, m=$N$) & 0.849 & \textbf{0.860} & - & \multirow{2}{*}{0.671} &  \multirow{2}{*}{0.819} \\
    Resize(s=$0.5$, m=$L$) & 0.908 & \textbf{0.920} & - &  & \\
    JPEG(q=$50$) & \textbf{0.831} & 0.749 & 0.67 & 0.817 & 0.746 \\
    \hline\noalign{\smallskip}
    \end{tabular}
    \end{center}
\end{table*}
\setlength{\tabcolsep}{1.4pt}

\subsection{Lossy compression versus watermark encoding}

Lossy compression algorithms and watermark encoders work in the same subdomain of the image, i.e., they try to modify pixel values that are not normally perceived, in order to reduce the size of the image or encode additional information in the image, respectively. Thus, we considered these algorithms as a special and sophisticated group of attacks. Assuming transparency of the watermark, the encoder should change those pixels that are removed or modified by the lossy compression algorithms. Therefore, in our work we mainly focus on preserving a robustness against lossy compression techniques as contemporary multimedia applications or services use them by default and it is impossible to skip the compression step due to technical  limitations of the broadcast bandwidth.
As  a result of the compression, the majority of the image space that is \textit{not perceivable} is removed, hence the watermarking method is not able to use this part of the image to encode the message (that would result in a perfect transparency). Consequently, we observe that the watermarking subtly affects the space-domain of the images that is \textit{perceived by humans} in order to retain the message after the compression. The effect of the compression algorithm is also observable in the case of the capacity. Note that in, somewhat similar to watermarking,  stenography, which does not typically consider attacks against the integrity of the message, we are able to embed in a cover image a message of the length of a separate image (or two)~\cite{Baluja,Baluja2}. In the case of the watermarking, due to the need of providing robustness against aimed attacks, we are able to handle only short messages (e.g. 30 bits~\cite{Zhu_2018_ECCV, luo2020distortion} or 1024 bits with severely constrained types of attacks~\cite{ReDMark}).      

\subsection{Robustness vs. quality of images}

\begin{figure*}
    \centering
    \includegraphics[width=0.95\textwidth]{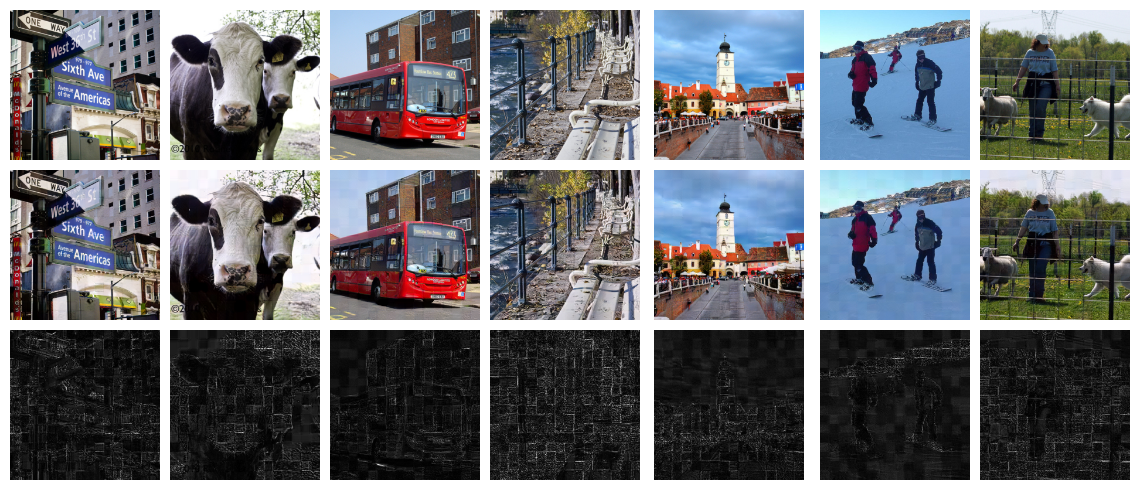}
    \caption{The comparison of the encoded image \(I_e\) (middle row) and the cover image \(I_c\) (top). The bottom row shows the min-max normalized difference between the cover image \(I_c\) and the encoded image \(I_e\). \newline Additional samples may be found at~\href{https://drive.google.com/drive/folders/1sqgAvXcanieYobqzFgt0tkLaVN20pII8}{https://drive.google.com/drive/folders/1sqgAvXcanieYobqzFgt0tkLaVN20pII8}}
    \label{fig:enc}
\end{figure*}

The method was evaluated for the PSNR equal to \(30.19\)dB, \(37.81\)dB and \(37.46\)dB for the Y, Cb and Cr channels, respectively. The quality of the encoded image is similar to results achieved in~\cite{Zhu_2018_ECCV}. In \cite{luo2020distortion,ReDMark}, authors reported slightly higher values of the PSNR. All methods achieved the quality of the images similar to the lossy compression algorithms \cite{barni2006document}, where the average PSNR for all channels is typically above  \(30\)dB. We did not take into consideration the results of robustness from~\cite{wen2019romark} because their method modifies the image significantly. In order to compare the distortion level we calculated the PSNR for our validation dataset after applying JPEG compression algorithm with the quality factor \(q = 50\) and the subsampling 4:2:0. We achieved the PSNR equal to \(36.35\)dB, \(36.78\)dB and \(36.92\)dB for the Y, Cb and Cr channels, respectively. And without using the subsampling technique, we achieved \(37.90\)dB, \(38.17\)dB and \(38.29\)dB. The results of the PSNR suggest that the message was encoded on the Y channel chiefly. The samples of encoded images were presented in Figure~\ref{fig:enc}\footnote{Additional samples can be found under the link: \href{https://drive.google.com/drive/folders/1sqgAvXcanieYobqzFgt0tkLaVN20pII8}{https://drive.google.com/drive/folders/1sqgAvXcanieYobqzFgt0tkLaVN20pII8}
}.

\section{Conclusions}

In the paper we propose a watermarking method based on spatial spreading of the message. Our architecture is done with convolutional neural networks and is scalable for any size of an image. We developed a special architecture for the encoder network, where the cover image and the message are yielded to every second layer. We also formulated a novel and custom loss function for training the neural networks. In comparison to previous method our watermarking system provides significantly improvement of robustness against Gaussian smoothing, resizing and JPEG (local attacks). The work is extended by additional attack types, such as subsampling 4:2:0 or rotation. We also achieve the bit accuracies above \(0.83\) for all considered attacks. This indicates that the method achieves high general robustness exceeding previous solutions. As a way to obtain our results, we conduced the experiments with grouping the attacks on the watermark based on their scope and we revealed some correlations between attacks. We show that in order to achieve the overall robustness of the watermarking method based on CNNs, we require to select an appropriate set of the attacks applied to the nosier layer. In future work we would like to continue to improve the robustness against the attacks, as well as apply and evaluate multi-attacks scenarios. We would like to increase the message capacity and extend the solution over a video domain and video-specific compression algorithms. Moreover, some other quality measures like the one presented in~\cite{Perceptual2016} may be considered in order to adjust the transparency. 

\textbf{Acknowledgments}. This work is partially supported by the National
Centre for Research and Development (NCBiR) Project
{POIR.01.01.01-00-1032/18} and Polish National Science
Centre (NCN) – Project {UMO-2018/29/B/ST6/02969}.

\bibliographystyle{unsrt}  

\bibliography{template}

\end{document}